\documentclass[aps,pra,twocolumn,showpacs,floatfix,amsmath]{revtex4}

\usepackage{graphicx}

\begin{document}
 \title{Elementary excitations of ultracold soft-core bosons across the superfluid-supersolid phase transition}
\author{T. Macr\`{i}, F. Maucher, F. Cinti and T. Pohl}
\affiliation{Max Planck Institute for the Physics of Complex Systems, N\"othnitzer Stra\ss e 38, 01187 Dresden, Germany}
\date{\today}

\begin{abstract}
We investigate the zero-temperature excitation spectrum of two-dimensional soft-core bosons for a wide range of parameters and across the phase transition from a superfluid to a supersolid state. Based on mean field calculations and recent Quantum Monte Carlo results, we demonstrate the applicability of the Bogoliubov-de Gennes equations, even at high interaction strengths where the system forms an insulating cluster crystal. Interestingly, our study reveals that the maximum energy of the longitudinal phonon band in the supersolid phase connects to the maxon energy of the superfluid at the phase transition.
\end{abstract}
\maketitle

A supersolid is a phase of matter that simultaneously accommodates diagonal as well as off-diagonal long-range order, which means that particles self-assemble into a rigid, regular crystal but at the same time can flow superfluidly through the formed solid. More than forty years ago, this peculiar state has been conjectured to emerge in pressurized solid Helium \cite{ali69,che70,leg70}, which led to an intense search for supersolidity in such systems \cite{mei92}. In 2004, experimental evidence for superfluidity in toroidal oscillator measurements \cite{kic04a,kic04b} has greatly revived interest in supersolid Helium and initiated a recent surge of activity on this problem. Yet, theoretical work has not reached a general consensus \cite{bkp06,bpb06,and09,bal10,bop12} as to whether solid $^4$He can display superfluidity, and very recent experiments \cite{kic12} are casting doubt on the original interpretation of the measurements \cite{kic04a,kic04b}.

At the same time, ultracold atomic gases have emerged as a promising alternative platform to realize and study continuous-space supersolids in an unambiguous and controlled fashion. Recent work has demonstrated that long-range soft-core interactions between bosonic atoms can be engineered via optical coupling to highly-lying electronic Rydberg states \cite{hnp10,pmb10,cjb10,hwp10,mhs11,hcp12,lhl12,grf12} or via light-induced interactions in an optical cavity \cite{mbb12}. Such soft-core interactions can give rise to so-called cluster-solids \cite{mcl08} and cluster-supersolids \cite{hnp10,cjb10}, i.e. crystalline arrangements of atomic clusters or droplets where superfluidity can arise from particle-hopping between the self-assembled droplets. 

On a more formal level, supersolidity can be understood in terms of the simultaneous breaking of fundamentally different symmetries: (\emph{i}) the breaking of translational symmetry responsible for the crystalline ordering and (\emph{ii}) the breaking a global gauge symmetry that enables long-range phase coherence and thereby superfluidity of the system. A direct consequence is the emergence of Goldstone bosons, i.e gapless modes in the excitation spectrum for each of the broken symmetries. Therefore, the excitation spectrum provides a powerful experimental way to probe supersolidity, e.g. via Bragg scattering \cite{skc99,sok02} and has recently attracted considerable theoretical interest \cite{smb12,wab12, kuk12,yej08,yod10,sjr08,sjr10}. Most recently, Quantum Monte Carlo \cite{smb11} simulations have been used to determine the dynamical structure factor of two-dimensional soft-core Bosons \cite{smb12}, while mean field approaches have been applied in numerous works \cite{pos94,jpr07a,jpr07b,wab12,kuk12,mjr12}, aiming at a simplified description of the zero-temperature physics of these systems. 

\begin{figure}[t!]
\resizebox{0.99\columnwidth}{!}{\includegraphics{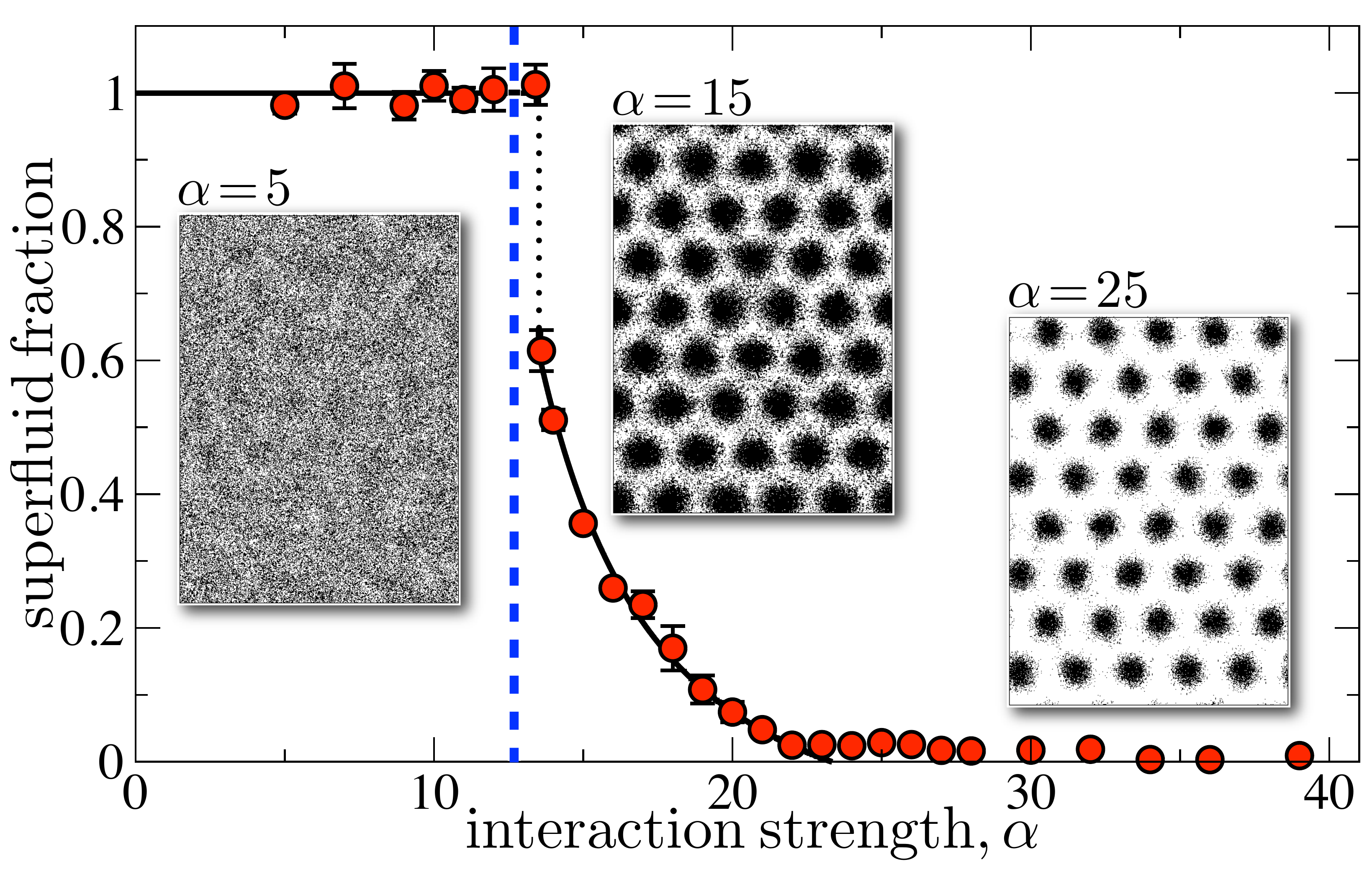}}
\caption{\label{fraction_s} Superfluid fraction of two dimensional Bosons as a function of the dimensionless interaction strength $\alpha=m\rho V_0 R_0^4/\hbar^2$. 
The points show results of Monte Carlo simulations for $320$ particles with a density of $R_0^2\rho=4.4$, whereas the continuous line is a guide for the eye. The superfluid fraction drops abruptly at $\alpha\approx13.4$ (dotted line), marking a first-order superfluid-supersolid phase transition, in close agreement with the mean field prediction of $\alpha=12.7$ (dashed line). Around $\alpha\approx22$ 
the systems enters an insulating phase. The insets show PIMC snapshots for different $\alpha$, illustrating the particle density profile in the three different phases. 
We checked that the results do not change for a temperature range $0.1-0.75\ \hbar^2/mR_0^2$ and particle number $160$, $320$, $640$.}
\end{figure} 

Here we present a thorough comparison between these two approaches and show that the excitation spectrum can be accurately described by the Bogoliubov-de Gennes (BdG) equations. Surprisingly, we find quantitative agreement with Monte Carlo simulations, not only in the superfluid and supersolid phases but also in the strongly interacting regime, where superfluidity is destroyed by quantum fluctuations and mean field approaches are generally expected to fail. This finding should prove valuable for future work, as it justifies the use of more efficient mean field calculations, which moreover enable investigations of dynamical processes, such as relaxation phenomena or externally driven systems. Here, we exploit this fact to study the excitation spectrum for a wide range of parameters, and find unexpected features at the superfluid-supersolid phase transition.

\section{Ground state properties} \label{GSPD}
We consider an ensemble of Bosons confined to two dimensions with with mass $m$ and positions ${\bf q}_i$, as described by the Hamiltonian
\begin{equation}\label{MC}
\hat H = \sum_i  -\frac{\hbar^2}{2m}  \nabla^2_i +
\sum_{i<j} V({\bf q}_i-{\bf q}_j).
\end{equation}
As a prototype example for soft-core interactions we chose a simple step function potential
$V({\bf r})=V_0 \Theta(R_0 - r)$, where $\Theta(r)$ denotes the Heaviside function and $V_0$ and $R_0$ define the strength and range of the interaction potential. Upon scaling lengths by $R_0$ and energies by $\hbar^2/m R_0^2$, the zero temperature physics, determined by eq.(\ref{MC}), depends only on two dimensionless parameters: an effective interaction strength $\alpha^{\prime}=V_0 m R_0^2/\hbar^2$ and the dimensionless density $\rho R_0^2$.
At the mean field level, this set of parameters can be further reduced. In this case, the system dynamics is described by a non-local Gross-Pitaevskii equation (GPE), which, in terms of the rescaled variables, can be written as
\begin{equation} 
\label{GPNH}
i \partial_t \psi_0({\bf r}) = \left[-\frac{ \nabla^2}{2} + \alpha \int {\rm d}{\bf r}^{\prime} U({\bf r}-{\bf r}^{\prime}) 
|\psi_0({\bf r}^{\prime})|^2\right] \psi_0({\bf r}) \;,
\end{equation}
where ${\bf r} = {\bf q}/R_0$, $U({\bf r})=\Theta(1-r)$, and $\alpha = m\rho V_0 R_0^4/\hbar^2$ is a dimensionless interaction strength that solely determines the dynamics and ground state properties. Such a mean field treatment is expected to be valid in the limit of weak interactions, reached by decreasing $\alpha^{\prime}$ and increasing the density $\rho$ such that the product $\alpha^{\prime} \rho$ stays finite. In this regime the zero temperature physics is determined only by the effective interaction strength $\alpha$, which greatly simplifies the analysis of the underlying phase diagram. 

For small $\alpha$ the system is in a homogenous superfluid phase whose energy $\pi\alpha$ follows directly from eq.(\ref{GPNH}). We describe the modulated supersolid state by a variational wave function that is composed of localized Gaussians, arranged on a triangular lattice. Their dispersion $\sigma$ and the lattice constant $a$ is obtained by minimizing the total energy. This simple analysis shows that density modulations become energetically favorable for $\alpha \ge 12.65$, marking a first order phase transition to a cluster supersolid state composed of small superfluid droplets \cite{hnp10,pmb10,smb11}. The droplets become more localized as the interaction strength increases where both their size as well as the lattice spacing decreases from $\sigma=0.39$ and $a=1.51$ at the phase transition to $\sigma=0.22$ and $a=1.40$ for $\alpha=40$.

\begin{figure}[t!]
\resizebox{0.99\columnwidth}{!}{\includegraphics{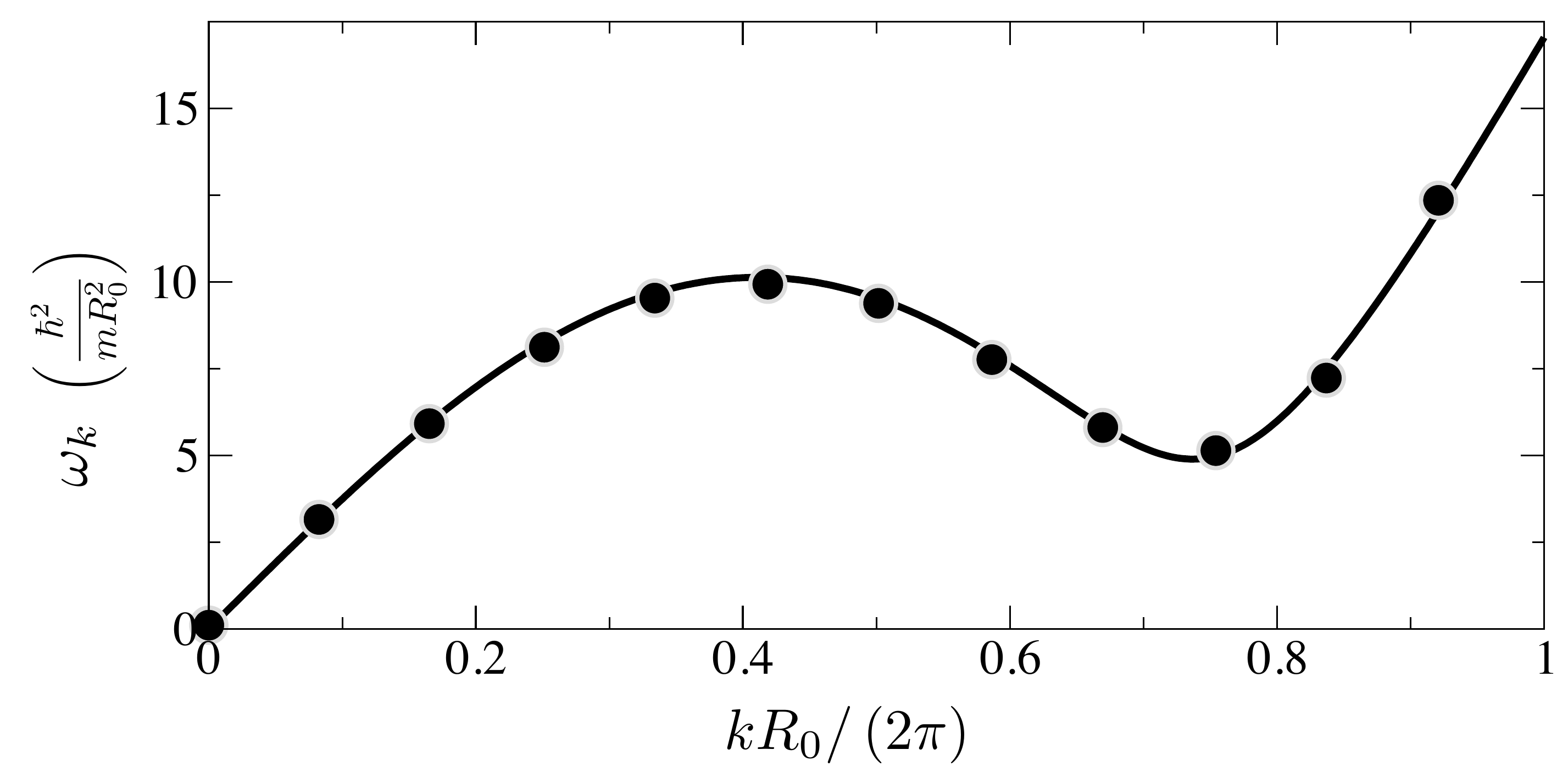}}
\caption{\label{fig2} Excitation spectrum in the superfluid phase according to eq.(\ref{bog}) (line) compared to the PIMC data (circles) of  Ref. \cite{smb12}.}
\end{figure} 

In order to test these predictions we additionally performed path integral Monte Carlo (PIMC) simulations at finite temperature based on the continuous-space Worm algorithm  \cite{bps06a,bps06b}, carefully extrapolating the zero temperature behavior. Fig.\ref{fraction_s} shows the obtained superfluid fraction as a function of $\alpha$ for a total number of $320$ particles with a density of $\rho R_0^2=4.4$ corresponding to eight particles per droplet. One finds a first order phase transition at $\alpha\approx13.4$, signaled by an abrupt drop of the superfluid fraction from unity in the homogenous superfluid phase to $\sim0.6$ at $\alpha=13.6$ in remarkable agreement with the above mean field prediction.
Also consistent with the above discussion, a further increase of $\alpha$ leads to a stronger localization of the droplets, and, hence, a drop of the superfluid fraction \cite{leg70}. Around $\alpha\approx22$ the particle density between the droplets decreases to a point where quantum fluctuations destroy phase coherence between individual clusters such that the system enters an insulating crystal of superfluid droplets without long range off diagonal order \cite{cjb10}. While this transition can evidently not be captured by eq.(\ref{GPNH}), the meanfield theory nevertheless yields an accurate description of the excitation spectrum in the insulating phase, as discussed below.

\begin{figure*}[t!]
\resizebox{0.99\textwidth}{!}{\includegraphics{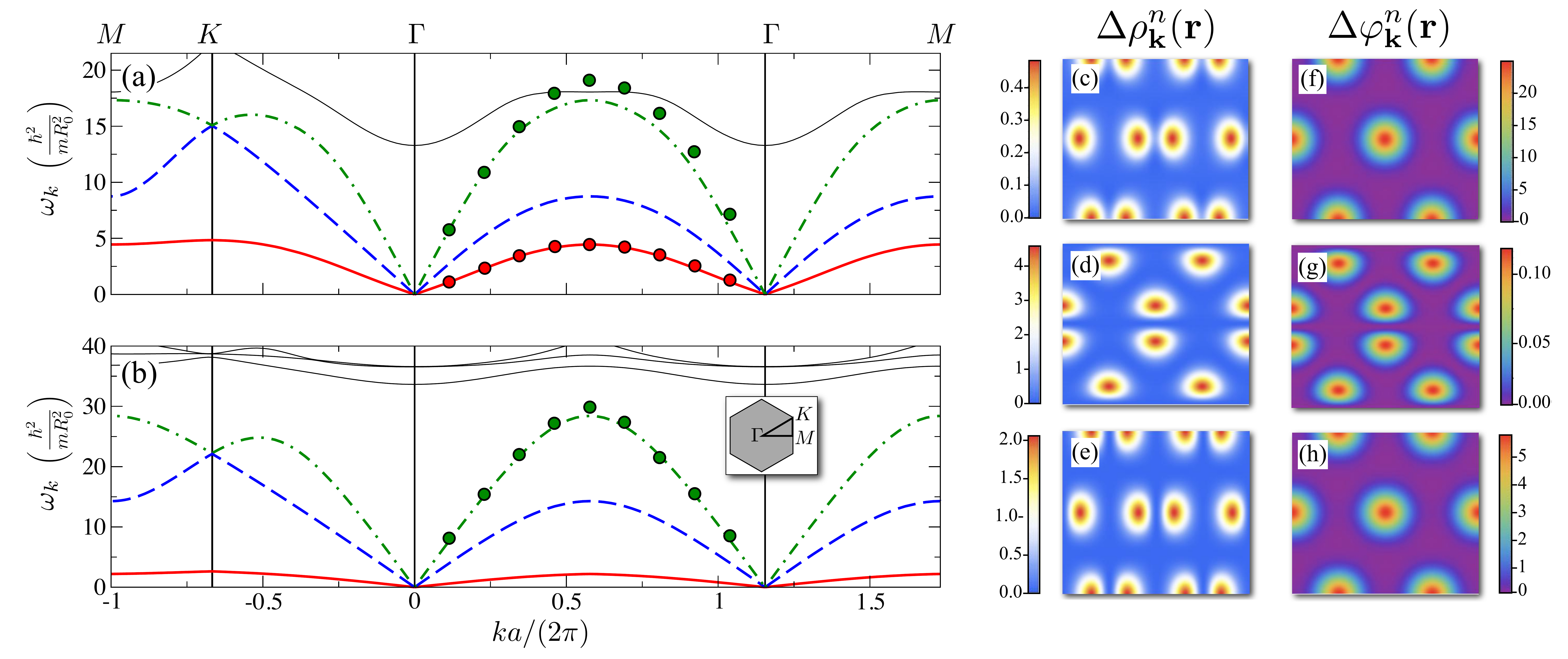}}
\caption{\label{MC_plot} 
(color online) Mean field spectra (lines) at $\alpha = 16.93$ (a) and $\alpha = 30.62$ (b) obtained from eqs.(\ref{BdGnuni}) along the three symmetry directions of the Brillouin zone [see inset of panel (b)]. The symbols represent the PIMC data of Ref. \cite{smb12} for  longitudinal excitations computed along the direction $\Gamma-M-\Gamma$ in the first two Brillouin zones. (c-e) Density fluctuations $\Delta \rho_\mathbf{k}^{n}(\mathbf{r})=\left| u_{n,\bf{k}}(\mathbf{r})-v_{n, \bf{k}}(\mathbf{r}) \right|^2$ and (f-h) phase fluctuations $\Delta \varphi_\mathbf{k}^{n}(\mathbf{r})=\left| u_{n,\bf{k}}(\mathbf{r})+v_{n, \bf{k}}(\mathbf{r}) \right|^2$ computed at $\alpha=16.93$ and $\mathbf{k}=\vec{\Gamma K}/10$ (directed along the horizontal axis) for the lowest (continuous line) band (c,f), middle (dashed) band (d,g), and higher (dot-dashed) gapless band (e,h).
}
\end{figure*}

\section{Excitations} \label{spectra}
The excitation spectrum is obtained by expanding the field, $\hat \psi=\psi_0+\delta\hat \psi$, around the groundstate. Substituting small perturbations of the form 
\begin{equation} \label{deltapsi}
\delta \hat \psi(\mathbf{r},t)=e^{-i\mu t}\sum_n \left[ u_n(\mathbf{r})e^{-i\omega t}a_n-v_n^*(\mathbf{r})e^{i\omega t}a_n^\dagger\right]\;,
\end{equation}
where $n$ labels the bands, into the GPE (\ref{GPNH}) yields to leading order in $\delta\hat\psi$ the familiar BdG equations for the Bogoliubov modes $u_n$ and $v_n$. We solve these equations in real space by expanding the modes into Bloch waves 
\begin{eqnarray} \label{bloch}
u_n({\bf r}) &=& u_{n,\bf k}({\bf r}) e^{i{\bf k}\cdot{\bf r}}\nonumber \\
v_n({\bf r}) &=& v_{n,\bf k}({\bf r}) e^{i{\bf k}\cdot {\bf r}}\;,
\end{eqnarray}
where the functions $u_{n, \bf k}({\bf r})$ and $v_{n, \bf k}({\bf r})$ obey the translational symmetry of the underlying groundstate, i.e. a continuous translational symmetry in the superfluid phase and a discrete triangular lattice periodicity in the supersolid phase. This ansatz leads to the following set of equations
\begin{widetext}
\begin{eqnarray}  \label{BdGnuni} 
\left( \frac{k^2}{2}-i\mathbf{k}\cdot\mathbf{\nabla} -\frac{1}{2}\nabla^2 +\alpha 
A({\bf r})-\mu \right)u_{n, \bf k}(\mathbf{r}) +
 \alpha \psi_0(\mathbf{r}) \int d\mathbf{r'}U(\mathbf{r-r'}) \psi_0(\mathbf{r'}) e^{i {\bf k}\cdot ({\bf r}^{\prime}-{\bf r})}
 [u_{n, \bf k}({\bf r}^{\prime})-v_{n, \bf k}({\bf r}^{\prime})] &=&\omega u_{n, \bf k}(\mathbf{r})\nonumber\\ 
-\left( \frac{k^2}{2}-i\mathbf{k}\cdot\mathbf{\nabla} -\frac{1}{2}\nabla^2 +\alpha 
A({\bf r})-\mu \right)v_{n, \bf k}(\mathbf{r}) +
 \alpha \psi_0(\mathbf{r}) \int d\mathbf{r'}U(\mathbf{r-r'}) \psi_0(\mathbf{r'}) e^{i {\bf k}\cdot ({\bf r}^{\prime}-{\bf r})}
 [u_{n, \bf k}({\bf r}^{\prime})-v_{n, \bf k}({\bf r}^{\prime})] &=&\omega v_{n, \bf k}(\mathbf{r})\nonumber\\
 & &
\end{eqnarray}
\end{widetext}
for $u_{n, \bf k}$ and $v_{n, \bf k}$, where $A({\bf r})=\int d\mathbf{r'}U(\mathbf{r'-r}) \left|\psi_0(\mathbf{r'})\right|^{2}$. Recently, the excitation spectrum has been investigated by calculating the dynamical structure factor via PIMC simulations \cite{smb12}, using the so-called GIFT approach \cite{vrr10}. In the following, these first-principle results will be employed to assess the predictive power of eqs.(\ref{BdGnuni}).

In the homogeneous superfluid phase, eqs.(\ref{BdGnuni}) can be solved analytically and yield the familiar Bogoliubov spectrum
\begin{equation} \label{bog}
\omega = \sqrt{\frac{k^2}{2}\left(\frac{k^2}{2}+2\alpha  \tilde{U}({\bf k})\right)},
\end{equation}
where $\tilde{U}({\bf k}) = 2\pi J_1(k)/k$ and $J_1$ denotes the Bessel function of the first kind.
For $\alpha \ge 5.03$ the spectrum develops a roton-maxon structure and roton softening occurs 
at $\alpha=14.74$, preceded by the supersolid phase transition at $\alpha=12.7$. Fig. \ref{fig2} illustrates the roton-maxon spectrum of the superfluid phase for an interaction strength of $\alpha=11.86$, i.e. slightly below the superfluid-supersolid phase transition. Eq.(\ref{bog}) is in excellent agreement with the numerical PIMC results of Ref. \cite{smb12}, indicating that the system can indeed be described as a weakly coupled fluid.

In the supersolid phase, a reliable calculation of the excitation spectrum requires accurate knowledge of the ground state  and its chemical potential. To this end we iterate the time-independent GPE \cite{sbe06}, $\mu\psi_0({\bf r})=[-\nabla^2/2+\alpha A({\bf r})]\psi_0({\bf r})$, starting from our optimized variational wave function and employing the same grid used to solve eqs.(\ref{BdGnuni}). 
In Fig.(\ref{MC_plot}) we show the obtained spectrum of low-energy excitations for two values of the interaction strength that lie in the supersolid ($\alpha=16.93$) and in the insulating crystal phase ($\alpha=30.62$), respectively. The figure shows the excitation energies along the three symmetry axes of the Brillouin zone corresponding to the underlying triangular lattice.
We find three gapless bands, i.e. three Goldstone modes reflecting the symmetries that are broken in the supersolid phase \cite{wab12}.
In addition to the "superfluid band" due to the breaking of global gauge symmetry, there are two bands corresponding to longitudinal 
and transverse phonon excitations of the two-dimensional lattice. While the latter were not accessible by the PIMC calculations of Ref. \cite{smb12}, we find good agreement for the two longitudinal modes in the supersolid phase ($\alpha=16.92$). Somewhat surprisingly, even in the insulating phase, eq.(\ref{BdGnuni}) yields excellent agreement for the longitudinal phonon mode, despite its evident inability to describe the break-down of global superfluidity. This indicates that each individual droplet maintains a high condensate fraction despite the apparent lack of global phase coherence between the crystalline ordered droplets (see Fig.\ref{fraction_s}).
A proper identification of each band can be done by computing local fluctuations on top of the mean field solution $\psi_0(r)$ \cite{wu96}. 
The substitution $\hat\psi(\mathbf{r})=e^{i\delta\hat\varphi(\mathbf{r})}\sqrt{|\psi_0(\mathbf{r})|^2+\delta\hat \rho(\mathbf{r})}$ allows us to identify local density and phase fluctuations:
\begin{equation} \label{OBDM}
\begin{array}{ccl}
\left<\delta \hat \rho^\dagger(\bf r) \delta\hat \rho(\bf r) \right>/\left|\psi_0(\mathbf{r}) \right|^2&=& \sum_{n,\bf k}
\left| u_{n,\bf{k}}(\mathbf{r})-v_{n, \bf{k}}(\mathbf{r}) \right|^2 \\
\left<\delta \hat\varphi^\dagger(\bf r) \delta\hat \varphi(\bf r) \right>\cdot 4\left|\psi_0(\mathbf{r}) \right|^2&=& \sum_{n,\bf k}
\left| u_{n,\bf{k}}(\mathbf{r})+v_{n, \bf{k}}(\mathbf{r}) \right|^2
\end{array}.
\end{equation}
Fig.\ref{MC_plot} (c-h) shows the contributions to (\ref{OBDM}) for one specific value of $\mathbf{k}$ for each of the three gapless bands at $\alpha=16.93$. One clearly distinguishes the transverse band from the direction of the fluctuations, orthogonal to the perturbing vector $\bf k$. The contribution of this band to phase fluctuations is strongly suppressed. The first and third band both contribute to density and phase fluctuations with different weight tough. The first band is mostly responsible for phase whereas the third to density fluctuations. Therefore the lower band can be associated to the superfluid response of the system, whereas the other two to the classical collective excitations of the crystal.

\begin{figure}[tT!]
\resizebox{0.99\columnwidth}{!}{\includegraphics{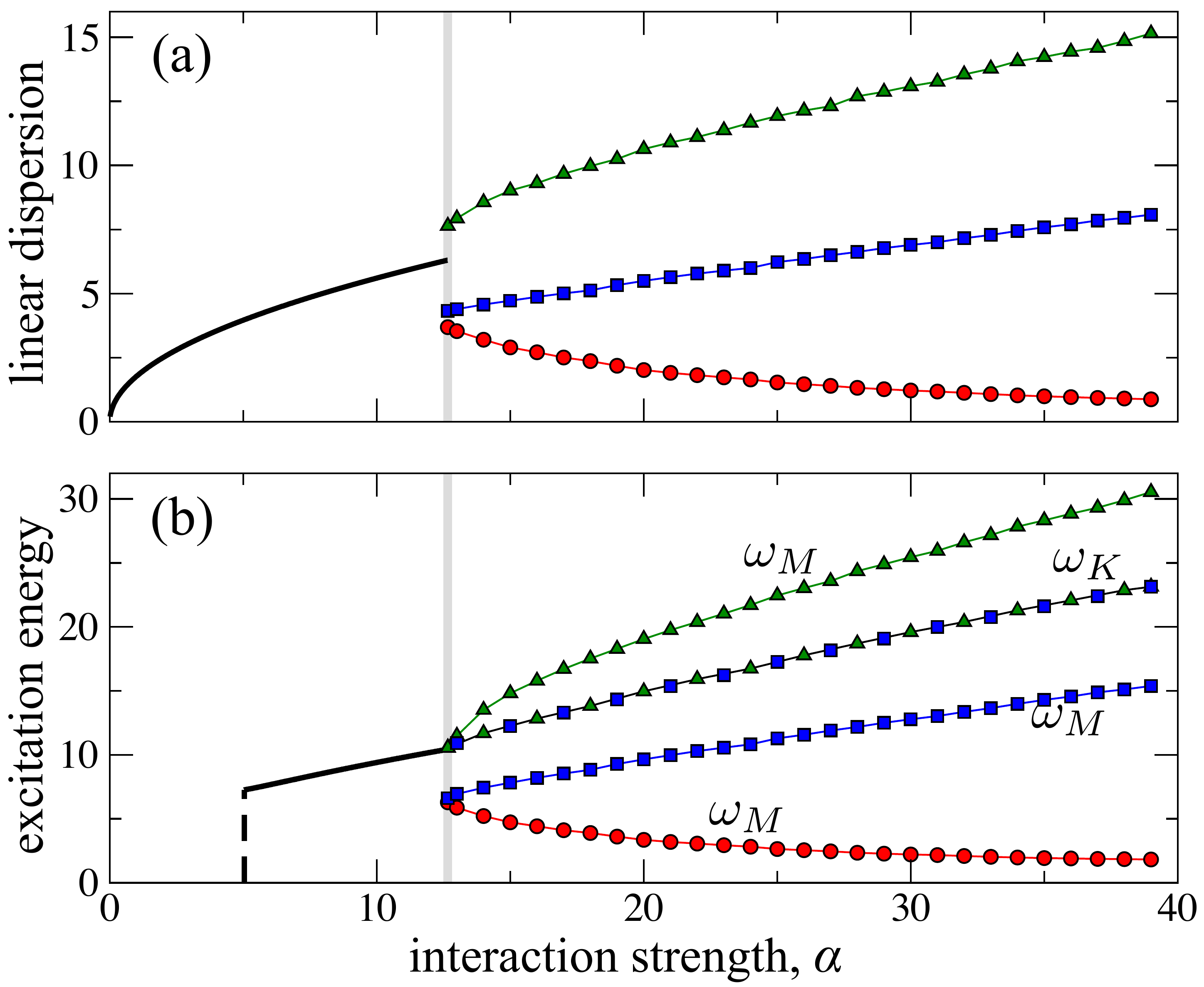}}
\caption{\label{velocity} (a) Linear dispersion $v=\partial \omega/\partial k|_{k=0} (\hbar^2/mR_0)$ of the gapless modes and (b) their energy $\omega (\hbar^2/mR_0^2)$ at distinct momenta as a function of $\alpha$. In the superfluid phase ($\alpha<12.7$) panel (b) shows the maxon energy (line) and the excitation energies (circles) of the three gapless modes at the $M$-point ($\omega_M$) and of the longitudinal and transverse phonon band at the $K$-point ($\omega_K$) in the supersolid phase (see Fig.\ref{MC_plot}).}
\end{figure}

Having demonstrated the accuracy of the mean field description we can now exploit its efficiency to study the mode structure over a wider range of interaction strengths and, in particular, across the superfluid-supersolid phase transition. The results are summarized in Fig.\ref{velocity}, which shows the linear dispersion $v=\partial \omega/\partial k|_{k=0}$ as well as the excitation frequency at several distinct momenta. As expected for a first order phase transition, the linear dispersion exhibits a discontinuous jump (Fig.\ref{velocity}a), reflecting the sudden onset of finite density modulations at the phase transition. On the other hand, the excitation energy shows a different and rather unexpected behavior. Fig.\ref{velocity}b shows the $\alpha$-dependence of the maxon energy in the superfluid phase along with the excitation energy of the three gapless bands at the $M$-symmetry-point of the Brillouin zone (see Fig.\ref{MC_plot}). At this symmetry point the longitudinal phonon mode provides the maximum energy in the entire reciprocal space. Interestingly, we observe that the maxon energy of the superfluid merges into the latter at the superfluid-supersolid phase transition, i.e. does not show a discontinuous jump. Remarkably, at the phase transition, this energy also coincides with the excitation energy at the other symmetry point ($K$-point) where, in addition, the longitudinal and transverse phonon modes are degenerate.

\section{Conclusion}
Based on PIMC simulations and meanfield calculations, we studied the zero-temperature physics of two-dimensional soft-core bosons and, in particular, their excitation spectrum at the phase transition to a supersolid droplet phase. 
The close match between both methods not only demonstrates the predictive power of the mean field approach but also attests to the accuracy of the GIFT method \cite{vrr10} for extracting dynamical properties from Quantum Monte Carlo simulations. Our spectra are consistent with recent calculations of Refs.~\cite{wab12,kuk12}, but differ qualitatively from the results of Ref.~\cite{yzl11}, where only one gapless mode has been found in the supersolid phase of soft-core dipoles. 

A careful scan of the particle interaction revealed that the maxon energy of the superfluid phase merges into the energy of longitudinal phonons at both reciprocal symmetry points of the supersolid phase. This degeneracy at the phase transition may suggest that the maxon part of the superfluid excitations plays a more significant role than thus far anticipated -- an implication which calls for further investigation. Future studies of other types of soft-core interactions \cite{hnp10,cjb10,mbb12} will clarify whether this behavior is of generic nature or a mere consequence of the step-function potential considered in this work. Moreover, PIMC simulations \cite{smb12} would allow to elucidate the effects of correlations, i.e. to address the question whether the found connection between maxon and phonon excitations persists in the strong coupling regime beyond the validity of the BdG eqs.(\ref{BdGnuni}). Along these lines, the behavior of Goldstone modes for related scenarios, e.g., at the crystallization point of dipolar systems \cite{dipole} or at the superfluid-solid transition of Helium \cite{bop12, pitaevskii92}, suggests itself as an interesting question for future studies. Finally, recent measurements that demonstrated supersolidity of ultracold atoms in an optical cavity via Bragg scattering \cite{mbb12} would provide a viable experimental way to investigate this question for yet another important type of soft-core interactions.

\section*{ACKNOWLEDGMENTS}
We thank S. Saccani, S. Moroni and M. Boninsegni for providing us their Monte Carlo data of Ref. \cite{smb12} and M. Boninsegni, G. Gori, N. Henkel, S. Moroni, L. P. Pitaevskii, A. Recati, S. Saccani, and A. Trombettoni for valuable discussions. This work was supported by the EU through the ITN COHERENCE.


\begin{thebibliography}{99}
\bibitem{ali69} A. F. Andreev and I. M. Lifshitz, Sov. Phys. JETP {\bf 29}, 1107 (1969).
\bibitem{che70} G. V. Chester, Phys. Rev. A {\bf 2}, 256 (1970).
\bibitem{leg70} A. J. Leggett, Phys. Rev. Lett. {\bf 25}, 2543 (1970).
\bibitem{mei92} M. W. Meisel, Physica B: Cond. Matt. {\bf 178}, 121 (1992).
\bibitem{kic04a} E. Kim and M. Chan, Nature (London) {\bf 427}, 225 (2004).
\bibitem{kic04b} E. Kim and M. Chan, Science {\bf 305}, 1941 (2004).
\bibitem{bkp06} M. Boninsegni et al., Phys. Rev. Lett. {\bf 97}, 080401 (2006).
\bibitem{bpb06} M. Boninsegni, N. Prokof'ev, and B. Svistunov, Phys. Rev. Lett. {\bf 96}, 105301 (2006).
\bibitem{and09} P. W. Anderson, Science {\bf 324}, 631 (2009).
\bibitem{bal10} S. Balibar, Nature {\bf 464}, 176 (2010).
\bibitem{bop12} M. Boninsegni and N. V. Prokof'ev, Rev. Mod. Phys. {\bf 84}, 759 (2012).
\bibitem{kic12} D. Y. Kim and M. H. W. Chan, Phys. Rev. Lett. {\bf 109}, 155301 (2012).
\bibitem{hnp10} N. Henkel, R. Nath and T. Pohl, Phys. Rev. Lett. {\bf 104}, 195302 (2010).
\bibitem{pmb10} G. Pupillo, A. Micheli, M. Boninsegni, I. Lesanovsky and P. Zoller, Phys. Rev. Lett. {\bf 104}, 223002 (2010).
\bibitem{cjb10} F. Cinti et al., Phys. Rev. Lett. {\bf 105}, 135301 (2010).
\bibitem{hwp10} J. Honer, H. Weimer, T. Pfau, and H. P. B\"uchler, Phys. Rev. Lett. {\bf 105}, 160404 (2010).
\bibitem{mhs11} F. Maucher et al., Phys. Rev. Lett. {\bf 106}, 170401 (2011)
\bibitem{hcp12} N. Henkel, F. Cinti, P. Jain, G. Pupillo, and T. Pohl, Phys. Rev. Lett. {\bf 108}, 265301 (2012).
\bibitem{lhl12} W. Li, L. Hamadeh and I. Lesanovsky, Phys. Rev. A {\bf 85}, 053615 (2012)
\bibitem{grf12} F. Grusdt and M. Fleischhauer, arXiv:1207.3716	
\bibitem{mbb12} R. Mottl et al., Science {\bf 336}, 1570 (2012).
\bibitem{mcl08} B. M. Mladek, P. Charbonneau, C. N. Likos, D. Frenkel and G. Kahl, J. Phys.: Condens. Matter 20, 494245 (2008).
\bibitem{skc99} D. M. Stamper-Kurn et al., Phys. Rev. Lett. {\bf 83}, 2876 (1999).
\bibitem{sok02} J. Steinhauer, R. Ozeri, N. Katz and N. Davidson, Phys. Rev. Lett. {\bf 88}, 120407 (2002).
\bibitem{smb12} S. Saccani, S. Moroni and M. Boninsegni, Phys. Rev. Lett. {\bf 108}, 175301 (2012).
\bibitem{wab12}  H. Watanabe and T. Brauner, Phys. Rev. D {\bf 85}, 085010 (2012).
\bibitem{kuk12} M. Kunimi and Y. Kato, Phys. Rev. B {\bf 86}, 060510 (2012).
\bibitem{yej08} J. Ye,  EPL {\bf 82}, 16001 (2008).
\bibitem{yod10} C.-D. Yoo and A. T. Dorsey, Phys. Rev. B {\bf 81}, 134518 (2010).
\bibitem{sjr08} N. Sepulveda, C. Josserand and S. Rica, Phys. Rev. B {\bf 77}, 054513 (2008).
\bibitem{sjr10} N. Sepulveda, C. Josserand and S. Rica, Eur. Phys. J. B {\bf 78}, 439447 (2010).
\bibitem{smb11} S. Saccani, S. Moroni and M. Boninsegni, Phys. Rev. B {\bf 83}, 092506 (2011).
\bibitem{pos94} Y. Pomeau and S. Rica, Phys. Rev. Lett. {\bf 72}, 2426 (1994).
\bibitem{jpr07a} C. Josserand, Y. Pomeau and S. Rica, Phys. Rev. Lett. {\bf 98}, 195301 (2007).
\bibitem{jpr07b} C. Josserand, Y. Pomeau and S. Rica, Euro. Phys. J. {\bf 146}, 47 (2007).
\bibitem{mjr12} P. Mason, C. Josserand and S. Rica, Phys. Rev. Lett. {\bf 109}, 045301 (2012).
\bibitem{bps06a} M. Boninsegni, N. Prokof'ev and B. Svistunov, Phys. Rev. Lett. {\bf 96}, 070601 (2006).
\bibitem{bps06b} M. Boninsegni, N. V. Prokof'ev and B. V. Svistunov, Phys. Rev. E {\bf 74}, 036701 (2006).
\bibitem{vrr10} E. Vitali, M. Rossi, L. Reatto, and D. E. Galli, Phys. Rev. B {\bf 82}, 174510 (2010).
\bibitem{sbe06} S. Skupin, O. Bang, D. Edmundson and W. Krolikowski, Phys. Rev. E {\bf 73}, 066603 (2006).
\bibitem{yzl11} X. Li, W. V. Liu and C. Lin, Phys. Rev. A {\bf 83}, 021602 (2011).
\bibitem{dipole} H. P. B\"uchler et al., Phys. Rev. Lett. {\bf 98}, 060404 (2007); G. E. Astrakharchik, J. Boronat, I. L. Kurbakov and Y. E. Lozovik, Phys. Rev. Lett. {\bf 98}, 060405 (2007).
\bibitem{wu96} Wen-Chin Wu and A. Griffin, Phys. Rev. A {\bf 54}, 4204 (1996).
\bibitem{pitaevskii92} L. P. Pitaevskii, J. of Low Temp. Phys. {\bf 87} 445 (1992).
\end{thebibliography}
\end{document}